\documentclass[10pt, a4paper, article]{IEEEtran}

\usepackage{float}
\usepackage{cite}

\usepackage{color}
\usepackage{lipsum}

\ifCLASSINFOpdf
  \usepackage[pdftex]{graphicx}
  \usepackage{epstopdf}
  \graphicspath{{./pics/}}
  \DeclareGraphicsExtensions{.pdf,.jpeg,.png}
\else
  \usepackage[dvips]{graphicx}
  \graphicspath{{./pics/}}
  \DeclareGraphicsExtensions{.eps}
\fi

\usepackage[cmex10]{amsmath}
\usepackage{amsfonts}
\usepackage{amssymb}

\ifCLASSOPTIONcompsoc
    \usepackage[tight,normalsize,sf,SF]{subfigure}
\else
    \usepackage[tight,footnotesize]{subfigure}
\fi

\usepackage{multirow} 
\usepackage{url} 

\usepackage{setspace} 
\usepackage{tikz}
\newcommand*\circled[1]{\tikz[baseline=(char.base)]{
            \node[shape=circle,draw,inner sep=0.5pt] (char) {#1};}}

\newcommand{\code}[1]{{\color{black}\texttt{#1}}} 

\begin{document}

\title{PIRATE: A Blockchain-based Secure Framework of Distributed Machine Learning in 5G Networks}
\author{Sicong~Zhou,
Huawei~Huang, Wuhui~Chen, Zibin~Zheng, and~Song~Guo

\thanks{S. Zhou, H. Huang, W. Chen and Z. Zheng are with the School of Data and Computer Science, Sun Yat-Sen University, China. Email: huanghw28@mail.sysu.edu.cn}

\thanks{Song Guo is with the Department of Computing, the Hong Kong Polytechnic University. Email: song.guo@polyu.edu.hk}
}

\maketitle

{\color{black}

\begin{abstract}
In the fifth-generation (5G) networks {\color{black} and the beyond}, communication latency and network bandwidth will be no more bottleneck to mobile users. Thus, almost every mobile device can participate in the distributed learning. That is, the availability issue of distributed learning can be eliminated. However, the model safety will become a challenge. This is because the distributed learning system is prone to suffering from byzantine attacks during the stages of updating model parameters and aggregating gradients amongst multiple learning participants. Therefore, to provide the byzantine-resilience for distributed learning in 5G era, this article proposes a secure computing framework based on the sharding-technique of blockchain, namely PIRATE. A case-study shows how the proposed PIRATE contributes to the distributed learning. Finally, we also envision some open issues and challenges based on the proposed byzantine-resilient learning framework.
\end{abstract}

\section{Introduction}\label{sec:Introduction}

{\color{black}

Machine learning has spawned a lot of useful applications, such as computer vision, and natural language processing, etc.
However, the parties who benefit from the technology are mostly large organizations, e.g., commercial companies and research institutes.
Individuals who only hold devices with limited computing-ability cannot take part in machine learning tasks.

In fact, the combined power of individuals by their holding devices has been much underestimated.
For instance, the mobility and the large population of the mobile users can bring flexible computing possibilities.
Taking these advantages into account, the distributed learning \cite{bertsekas1989parallel} enables individual devices to learn collaboratively.
However, device heterogeneity and network conditions have been viewed as the challenges of distributed learning.

On one hand, a recent literature \cite{Lian2017async} shows that asynchronous distributed learning starts to have a sound theoretical basis.
In the synchronous manner of distributed-learning, the unequal training times in each iteration induce the inevitable waiting dissipation on the fast training nodes. This problem brought by device heterogeneity will be eliminated with the growing maturity of the asynchronous learning.

On the other hand, communication latency and bandwidth are still viewed as the bottleneck resources of distributed machine learning \cite{lian2017can}.
Nowadays, the network bandwidth of most users is still insufficient to support the distributed-learning tasks.
This situation makes distributed learning highly unavailable for the majority.
Fortunately, latency and bandwidth will be no more obstacles to distributed learning in the era of the fifth-generation (5G) networks, because the network conditions of individuals will be substantially improved.

Generally, the ideal factors that affect the distributed learning include the following aspects:

\begin{itemize}
    \item High availability: any device can perform learning anytime.
    
    \item High scalability: the learning framework should support high concurrency, high communication efficiency and low storage complexity. 
    
    \item Decentralization: the intervention of a centralized third-party is minimum.
    
    \item Byzantine-resilient model safety: the future distributed learning should ensure byzantine-resiliency \cite{lamport}, which indicates that the distributed learning can endure arbitrary attacks.
\end{itemize}

Since the availability of  distributed-learning systems can be significantly improved in the 5G  networks, we can reasonably envision a distributed learning system in which the heterogeneous devices are available anytime. This is because the 5G technologies enable various devices to participate in distributed-learning tasks with the sufficient network bandwidth.
In such a sound distributed learning system, another challenge is that the learning system is prone to suffering from \emph{byzantine attacks} \cite{lamport}.
Therefore, more sophisticated approaches that can ensure the byzantine-resilience are required.
}

{\color{black}
Recently, byzantine-resilient machine learning under master/slave settings has gained much attention \cite{KRUM, chen2017distributed, li2019abnormal}. 
In a byzantine-resilient distributed-learning task, two things are risk-prone: 1) gradient aggregation, and 2) model parameters.

{\color{black}
Conventionally, the byzantine-resilient distributed-learning tasks are conducted under centralized settings, in which the byzantine-tolerant components rely on a globally trusted third-party as the \emph{parameter server}.
}
The problem is that the workload-handling capacity of such the centralized parameter server is usually a bottleneck while performing the distributed learning.
Moreover, to provide reliable services against the vulnerability of single-point-of-failure (SPOF), the redundant deployment of resources is entailed at the centralized third-party. Therefore, the centralized byzantine-resilient learning induces a high operational-expenditure (OPEX).

 To achieve high availability while fulfilling the requirement of OPEX-efficiency and byzantine-resiliency, the decentralized learning system would ideally be decentralized.
 {\color{black}For example,} a recent study \cite{lian2017can} {\color{black}claims} that the decentralized learning system is more efficient in terms of communication and storage {\color{black}than the centralized learning systems}.
 {\color{black}However, to the best of our knowledge,} the byzantine-resilient learning in a decentralized manner has not been well studied. 
}

{\color{black}
{\color{black}Although the} decentralized scheme shows the promising efficiency, a great challenge in {\color{black}terms of model-}safety comes along with it. That is, the aggregation {\color{black}stage of distributed learning} can be easily falsified, since the {\color{black}distributed-learning} task becomes a collaboration among multiple peers. Once the model parameters are contaminated at one of the peers, this local-model contributor would bring limited {\color{black}or even negative} contribution to the holistic learning task.
Thus, to achieve the decentralized byzantine-resilience for distributed learning, a \emph{state machine replication} (SMR) protocol \cite{lamport}  is require.
{\color{black}As a representative paradigm of SMR protocols, the blockchain technology can provide such the critical byzantine-resilience to distributed learning.}
In detail, blockchains can provide protection to the model parameters, particularly at the stage of gradient aggregation, while ensuring a high level of availability, cost-efficiency and decentralization.

To this end, this article proposes a sharding-based blockchain framework, which is named PIRATE, for byzantine-resilient distributed-learning under the decentralized 5G computing environment.

}

\begin{figure*}[t]
  \centering
  \includegraphics[width=1\linewidth]{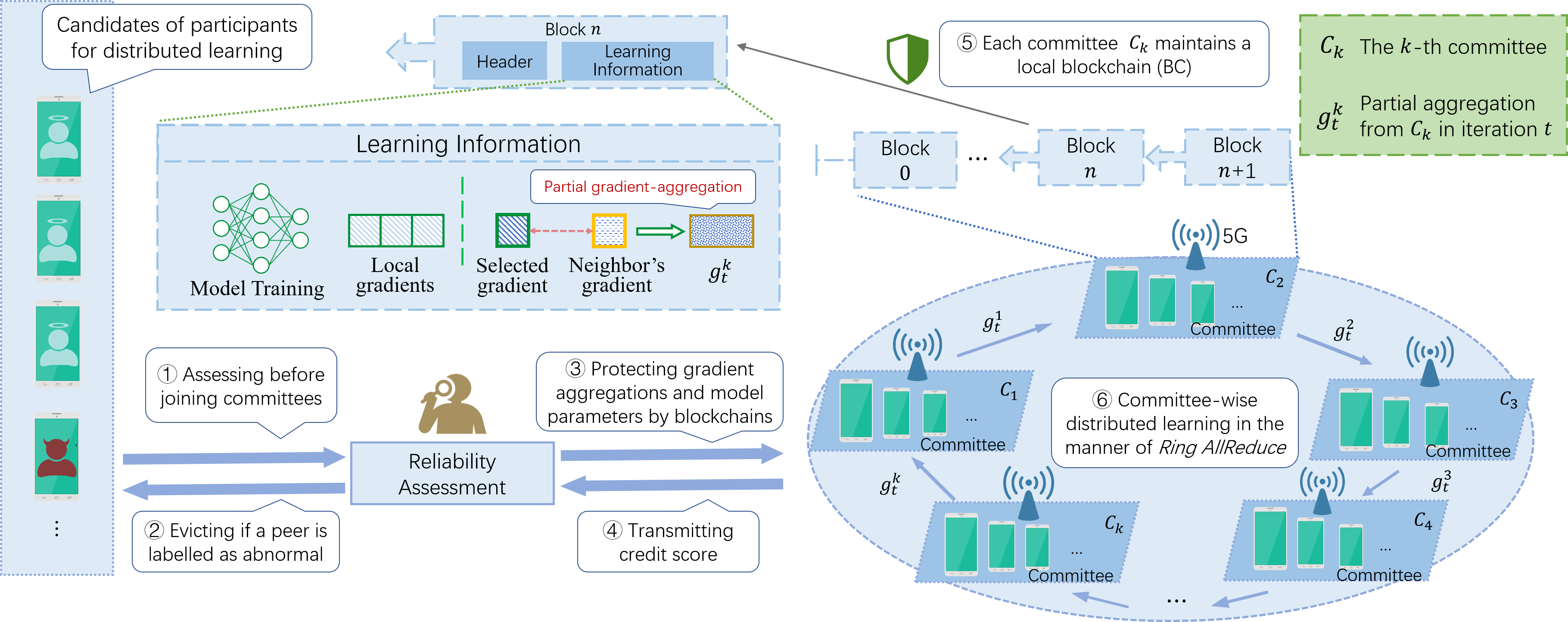}
  \caption{The proposed PIRATE framework has two critical components: 1) reliability assessment, and 2) blockchain (BC) systems for distributed SGD (D-SGD). Gradient aggregations and model parameters are protected by blockchains. Meanwhile, reliability assessment determines the participants of distributed learning tasks.}\label{fig:world}
\end{figure*}

{\color{black}
\section{Preliminaries of Distributed Machine Learning in 5G Networks}\label{sec:DML-5G}
}


\subsection{Consensus Protocols for Decentralized Learning in 5G}

{\color{black}

To achieve the global agreement within a decentralized setting, a byzantine-tolerant consensus protocol for SMR is needed.
Such the consensus protocol should ensure that the honest nodes can reach an agreement on the order of correct execution, i.e., model updates, even when a certain amount of byzantine effort plays a role \cite{lamport}. 
In this section, we first briefly review the existing byzantine tolerant consensus protocols towards SMR, and then analyze what protocol is applicable for decentralized learning in the 5G era.
For brevity, we call the byzantine-tolerant consensus protocol the consensus protocol in the remainder of this paper.

}

{\color{black}

Consensus protocols can be categorized into 2 types: {\color{black}the competition-based} and {\color{black}the communication-based}.
The leader of competition-based consensus, e.g., the Proof of Work (PoW) {\color{black} adopted by bitcoin (\url{https://bitcoin.org})}, needs to earn his leadership through a ``fair" competition. The communication-based consensus protocols (e.g., {\color{black} Hotstuff} \cite{yin2018hotstuff}) select leaders through a deterministic way, or based on an unbiased randomness generated collaboratively.

}

{\color{black}

{\color{black}In the context of blockchains,} for the competition-based consensus protocols, blocks are appended on the chain \emph{before} consensus while the communication-based protocols append blocks thereafter.
}
For the competition-based methods, higher scalability inevitably incurs higher chance of forking. Plagued by the byproduct, competition-based methods struggle to achieve a high scalability.

While communication-based consensus protocols have no concern of forks (instant finality), they require multiple rounds of communication to reach an agreement. The high communication overhead also hurts the scalability of communication-based protocols. Therefore, the communication-based protocols are commonly used at a smaller scale.

Sharding-based consensus protocols {\color{black}\cite{zamani2018rapidchain}}, as a hybrid approach, can achieve scalable consensus in the permissionless blockchain. Such hybridization benefits from the instant finality of the communication-based methods, and permissionless resiliency of the competition-based methods.
One representative work of the sharding-based approaches is \emph{RapidChain} \cite{zamani2018rapidchain}, which achieves a throughput of 7,300 tx/sec while maintaining a total resiliency of 1/3, i.e., the RapidChain system functions properly in face of 1/3 byzantine-malicious participants.

{\color{black}

In 5G networks, we need a consensus protocol that is available for a large scale of participants when they are training large models.
An ideal consensus protocol should be scalable and permissionless.
Sharding-based protocols make full use of integrated resources by splitting a workload and amortizing all tasks among multiple committees. 
Such scalable strategy can fit perfectly in the environment of distributed machine learning.
 However, permissionless distributed-learning in a sharding-based community is challenging.
 Unlike a financial system where good behaviors can be stimulated by a certain incentive mechanism, when performing the distributed-learning tasks, only the honest behavior is not enough.
 Because participants cannot adjust the network conditions and computation abilities of their devices according to their behaviors. 
 Thus, a consistent assessment of training-reliability throughout the training process is required for the distributed-learning task to function efficiently at each participant.
  Accordingly, we adopt a permissioned version of sharding-based consensus protocol in our proposed framework.

}

\subsection{Configurations of Distributed Machine Learning}\label{subsec:DML}

{\color{black}

Machine Learning problems mostly rely on certain specific optimization problems.
In the distributed machine learning, each participant can only yield a partial solution.
 In order to orchestrate a global optimization, as shown in Fig. \ref{fig:world}, computing nodes need to carry out the following steps.
 First, each node computes local gradients based on their local data. 
 Then, nodes communicate with each other to get a globally aggregated gradient for model update. 
  Based on the style how they get a global aggregation, two typical styles of configuration have been proposed:

}

{\color{black}

  \emph{Master/slave} style: A centralized parameter server aggregates gradients and synchronizes the aggregated result at each computing node.

  \emph{Decentralized} style: No centralized server is entailed. Each node can only communicate with its neighbors to exchange model parameters.
  
 }

{\color{black}

 We then discuss the superiority of decentralization on distributed machine learning:
 
 \begin{itemize}
    \item \textbf{Communication efficiency}. A recent work \cite{lian2017can} demonstrated that the decentralized settings can better exploit the bandwidth resource, avoid traffic jams and share workloads among peers than the centralized master/slave settings.

     \item \textbf{Cost efficiency}. When the scale of participation grows substantially, no single party should be responsible for maintaining the system.
     Analogous to the situation of cloud netdisk services nowadays, if a service provider plays such a dominant role, the OPEX cost of maintaining the centralized system eventually transfer to clients.
     As a feasible solution, service providers enforce clients to choose between low quality of service (e.g., very limited download speed) and expensive membership fee.
     This is an opposite of win-win for both providers and clients, provided that there are better solutions. 
     
     \item \textbf{Reliability}. The centralized design of master/lave settings suffer from SPOF problem. Once the centralized server is overloaded or under attack, the whole system ceases to function. Thus, the communication burdens and the attack risks on one single server impair the reliability of the system.
     
 \end{itemize}

}

{\color{black}

\section{State-of-the-art studies of Byzantine-Resilient Machine Learning}\label{sec:RelatedWork}

}

\subsection{Byzantine-Resilient Machine Learning}

{\color{black}

 To train models in a large-scale multi-party network, serious security problems are needed to address if adopt the Sharing technique. 
 As the scale of participants grows, the behaviors of computing nodes become more unpredictable. 
 The distributed stochastic gradient descent (D-SGD)  \cite{zinkevich2010parallelized} framework should tolerate byzantine attacks, i.e., arbitrary malicious actions, such as sending harmful gradients and corrupting training models as shown in Fig. \ref{fig:ring}.
 Current studies on Byzantine-resilient machine learning mostly focus on protecting gradient aggregation. However, the parameters of training models owned by each data trainer are also vulnerable to be falsified.
 We elaborate the protection of gradient aggregation in the next section, and analyze existing frameworks that can protect both aspects of model training.

}

{\color{black}

 A blockchain-based learning method was proposed by Chen \emph{et al.} \cite{chenLearningChain}, in which the proposed architecture \emph{LearningChain} is able to simultaneously protect gradient aggregation and model parameters, by storing them together on-chain.
 By exploiting the traceability of blockchain, erroneous global parameters produced by malicious nodes can be rolled back to its former unfalsified state. Historical parameter records cannot be falsified due to the tamper-proof character of blockchain.
 The proposed  byzantine-tolerant gradient aggregation with a low computation-complexity is called ``\emph{l}-nearest gradients aggregation". This method ensures that if byzantine computing nodes yield local malicious gradients to prevent the convergence of learning algorithms, their effort would be mitigated.
 However, \emph{LearningChain} still utilizes a master/slave setting for distributed gradient descend where the parameter server is randomly selected by PoW competition.
 In addition, the on-chain data could be potentially oversized, because any node in the system would have to store all the historical model parameters and gradients. Such architecture is prone to lead to a significant traffic congestion and a substantial storage overhead.
 When the scale of learning grows substantially, the storage and communication overheads could potentially weaken the advantage of distributed learning. 

}

{\color{black}

 In terms of reliability, rollback  can possibly fail if two consecutive byzantine leaders collude. 
 That is the case when the previous byzantine leader contaminates the model updates and the subsequent leader properly updates but based on the contaminated model parameters.
 The honest leader following the two previous byzantine leaders is unable to detect the model contamination, since the model-examination process terminates when the honest leader approves the immediate update proposal, even though the update is based on contaminated model parameters.
  The essential problem is that, the model update is examined by only one leader when a proposal is submitted.
  In contrast, our proposed framework adopts a more decentralized setting, in which all nodes can naturally participate in validating the model updates, and every node maintains its own training model rather than accepting updates from a third party. 
 
 }

\subsection{Byzantine Protection on Gradients}

{\color{black}

 Before the stage of updating training models, computing nodes need to aggregate their local gradients. The conventional aggregation solution of simple linear combinations {\color{black}(e.g., averaging \cite{polyak1992acceleration})} cannot tolerate one byzantine worker \cite{KRUM}. Thus, byzantine protection on gradients aggregation has gained much growing attention. Basically, the existing byzantine-based approaches can be classified into 2 categories: the tolerance-based methods and the detection-based methods.

}

\begin{table*}[t]
\centering
\caption{Gradient protection methods}\label{Tab:aggregator_compare}
\begin{tabular}{|l|c|c|c|c|c|}

\hline

     \code{Method types} & \multicolumn{2}{|c|}{\code{The tolerance-based}}     & {\code{The detection-based}} & {\code{The learning-based}}    \\ \hline
     
    \multirow{2}*{\code{Representative studies}} & \multirow{2}*{Krum \cite{KRUM}}                        & {$l$-nearest} & \multirow{2}*{Anomaly Detection \cite{li2019abnormal}} & \multirow{2}*{ Learning to learn\cite{L2L}}   \\
    
    {} & {} & {gradients \cite{chenLearningChain}} & {} &  {} \\ \hline

    \multicolumn{1}{|l|}{Resiliency under 30\% attack} & High & {Medium} & {Unknown}              & {High} \\ \cline{1-1}
    \multicolumn{1}{|l|}{Resiliency under 30\% attack (FL)} & Low & {Unknown}              & {High} & {Unknown}  \\ \cline{1-1}
    \multicolumn{1}{|l|}{Resiliency under majority attack} & Low & {Medium} & {Unknown} & {High}             \\ \cline{1-1}
    \multicolumn{1}{|l|}{Resiliency under majority attack (FL)} & Low & {Unknown} & {High} & {Unknown}             \\ \cline{1-1}
    \multicolumn{1}{|l|}{Computation complexity}         & O($n^2$) & {O($n$)} & {O($n$)} & {Model-related}       \\ \hline
   
    \multicolumn{1}{|l|}{Other functions except \emph{aggregation}}                & \multicolumn{2}{|c|}{None}                         & {Autoencoder training} & {Aggregator training} \\ \hline
    \end{tabular}

\end{table*}

{\color{black}

 Blanchard \emph{et al.} \cite{KRUM} proposed a byzantine tolerant method called \emph{Krum} , and a variation called \emph{Multi-Krum} which is an interpolation between Krum and averaging \cite{polyak1992acceleration}. 
 Instead of using a straight-forward linear combination of multiple local gradients, Krum precludes gradients too far away from the majority and chooses one local gradient based on a spatial score.
 To improve convergence speed, Multi-Krum chooses a number of local gradients based on the same score as Krum, and outputs their average. Experiments show that even with 33\% of omniscient byzantine workers, the error rate is almost the same as that of 0\%. However, Krum’s computation complexity is O($n^2$). Compare with the averaging approach whose computation complexity is O($n$), Krum may impede the scalability of distributed learning.
 }

{\color{black}

 The heuristic byzantine tolerant aggregation solution \emph{l}-nearest gradients proposed by Chen \emph{et al.} \cite{chenLearningChain} cannot guarantee safety against omniscient attacks by byzantine workers. The aggregation solution is to aggregate \emph{l} gradients closest, based on their cosine distances, to the sum of the received gradients.
 If an omniscient byzantine worker manages to acquire all local gradients for other workers in time, the byzantine worker can yield a local gradient that changes the global sum arbitrarily \cite{KRUM}. Sacrificing security, the algorithm gains a good time complexity O($n$).

}

{\color{black}

 One byzantine-detection method proposed by Li \emph{et al.} \cite{li2019abnormal} is designed for federated learning (FL). Existing byzantine-tolerant aggregation methods are mostly inefficient due to the non-identically and independently distributed training data. Experiments show that their detection-based method has a better performance than tolerance-based methods in federated learning.
 In their setting, a credit score was assigned by the pre-trained anomaly-detection model to each computing node. Since the weight of the local gradient was determined by the credit score, the weighted sum aggregation can filter out the byzantine local gradients.
 }

{\color{black}

 Besides the tolerance and detection methods, another machine learning method was proposed by Ji \emph{et al.} \cite{L2L} to learn the gradient aggregation. Different from the deterministic aggregation solution, they model the aggregation as a learning problem. They utilize an Recurrent Neural Network (RNN) aggregator with all local gradients as input, and additional loss information, i.e., the loss based on a small sample of dataset, for each worker. The RNN aggregator optimizes an objective that depends on the trajectory of the original optimization problem.
 }

{\color{black}

 The tolerance-based gradient aggregation methods are mostly designed under an independent identically distributed (i.i.d) assumption. Therefore, in the settings of federated learning where data is non-i.i.d, most tolerance-based methods do not perform well. The relaxation of data distribution brings challenge to the tolerance-based methods.
 However, the tolerance-based methods have the benefit of simplicity that do not require additional training. Non-i.i.d tolerance-based methods are challenging and interesting at the same time.
 }

{\color{black}

 As shown in Table. \ref{Tab:aggregator_compare}, we compare the performance of different  protection approaches for gradient-aggregation, and analyze their computation complexities, under two settings, i.e., the normal setting and the federated learning setting.

}

\subsection{Protection over Decentralized Machine Learning}

{\color{black}

 Recall that we have analyzed how a decentralized learning system is more promising than the centralized in terms of efficiency and reliability at large scale networks. However, the decentralized byzantine-resilient aggregation and the parameter protection for byzantine models have been ignore. 
 }

{\color{black}

 In a decentralized scheme, each node, including the malicious one, has a great impact on the global aggregation. As shown in Fig. \ref{fig:ring}, the existing protection methods are not applicable to the decentralized settings because of the following reasons.
 
 \begin{itemize}
    \item Attacks on partial aggregation are detrimental. Every node aggregates a partial aggregation provided by another node. For byzantine nodes, they have more attack patterns undermining learning tasks, such as sending falsified partial aggregation results or sending nothing to interrupt the aggregation process.
    Thus, without a quorum of validators in a multi-party network, partial aggregation process cannot be trusted.
    
    \item Without a trusted third party, anomaly-detection would be challenging. Credit scores of learning can be easily falsified without proper validation mechanisms.
    
    \item Moreover, synchronization of model parameters could be a problem. Every node needs to maintain a local training model. Once contaminated by attackers, computing nodes will have no actual contribution to the holistic learning task.
 \end{itemize}
}

{\color{black}

 In summary, blockchain as a decentralized SMR system, can provide a quorums of validators and practical synchronization mechanism to achieve byzantine-resiliency in decentralized learning.
 }

\section{Our Proposal - PIRATE: a Machine Learning Framework based on Sharding Technique}

\subsection{Overview of PIRATE}

{\color{black}

 Generally, we propose a framework of blockchain-based protection framework for the distributed-machine learning that is named PIRATE and managed in a decentralized manner.
 To protect both the gradient aggregation and model parameters, the proposed approach utilizes a sharding-based blockchain protocol and gradient anomaly-detection \cite{li2019abnormal}.
 Furthermore, with anomaly detection, the malicious nodes who yield harmful gradients can be replaced by eliminating their effort.

 The proposed framework can be also applicable for federated learning under a decentralized configuration, where no a centralized parameter server is required.

}


\subsection{Permission/Access Control}

{\color{black}

 In a distributed learning task with high availability, reliability assessment is essential for all participants, especially the mobile devices, which have volatile states.
 For instance, devices could be busy running some CPU-consumed tasks or simply charging in a sleeping mode. In the perspective of network conditions, they can be well connected in a net cafe or without signal at all while travelling through a tunnel. Allowing devices that are in bad states to keep their learning tasks would severely slow down the entire learning process. Thus, the volatile states of devices demand real-time reliability assessment to conduct permission control.
 
 }

{\color{black}

 We propose a centralized solution for permission control.  Fig. \ref{fig:world} depicts the permission/access control of PIRATE. Similar to federated learning, before actually contributing to the global learning task, all nodes are assessed by a centralized component based on their computation ability, network condition, join/leave prospect and historical credit scores. Nodes once granted permission, can join a committee using the Bounded Cuckoo rule \cite{zamani2018rapidchain}.

 During training, validated credit scores generated by committees are transmitted to the permission-control center. Nodes with low accumulated credit scores would be evicted from the system.

 }

\begin{figure}[t]
  \centering
  \includegraphics[width=1\linewidth]{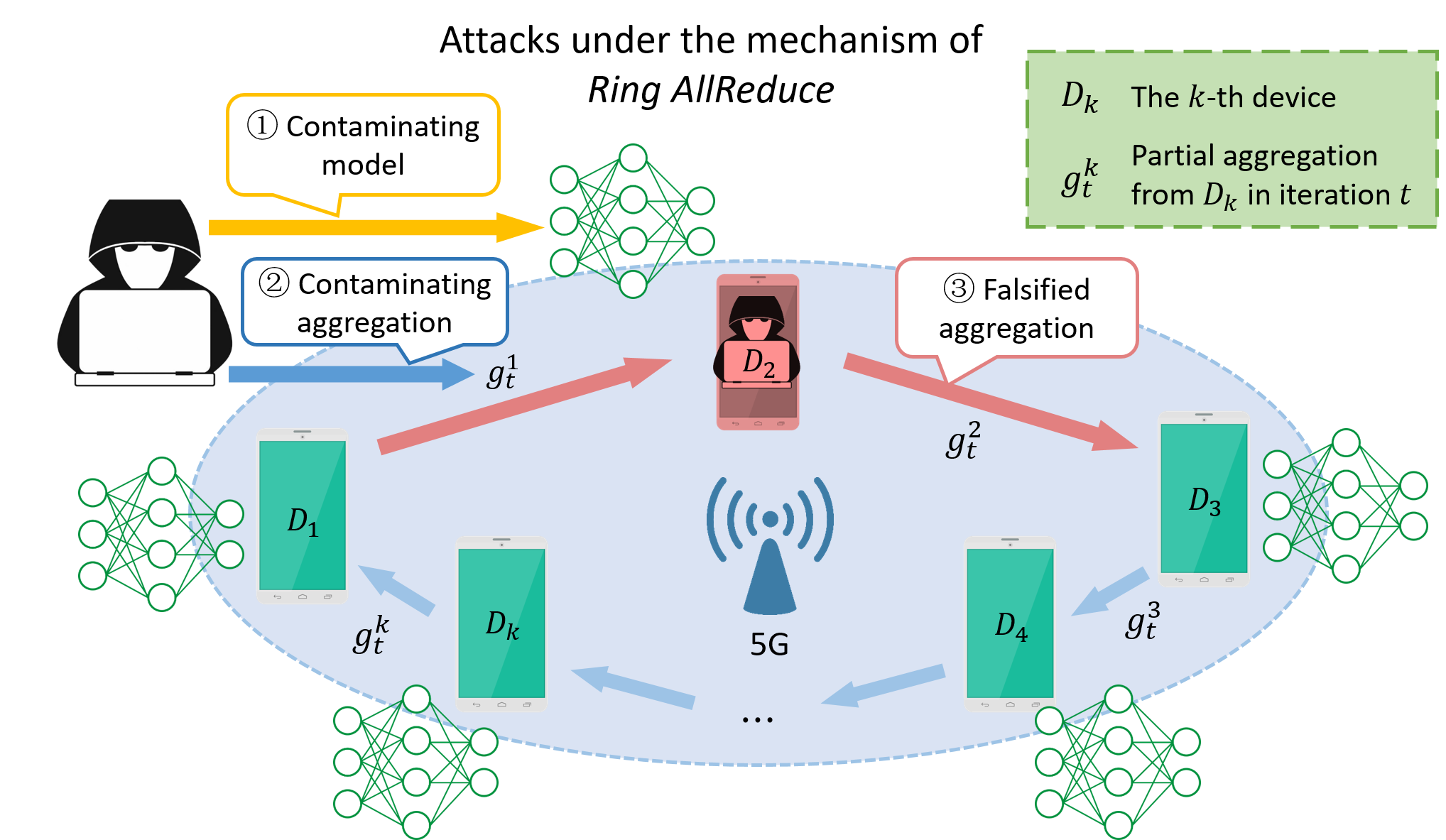}
  \caption{ 
   While adopting the \emph{Ring Allreduce}{\protect \footnotemark[1]} mechanism, malicious attackers can perform attacking from both inside and outside.
   \protect \circled{1} Attackers from the outside can contaminate training models in target nodes. \protect \circled{2} The outside attackers can also attack partial gradient-aggregation. \protect \circled{3} Byzantine computing nodes can send harmful aggregations that damage the convergence of learning tasks.}\label{fig:ring}
\end{figure}


\subsection{Sharding-based Blockchain Protection towards Decentralized Distributed-Learning}

{\color{black}

 We propose a sharding-based blockchain mechanism for the protection of decentralized distributed-learning. Existing byzantine-resilient aggregation solutions are not applicable under the decentralized settings. Having considered the increased freedom of byzantine nodes, our mechanism provides a consensus protocol on model parameters and gradient aggregation. Taking the advantages of 5G techniques, the sharding-based protection can work efficiently.
 }

 We randomly split the computing nodes into multiple committees, in which partial aggregations are agreed among committee members. Since aggregations are no longer centralized, the burden of aggregation workload is mitigated.

 Let $n$ denote the total number of computing nodes, $c$ denote the size of a committee, we assume that through permission management, every node’s computation time for local gradients is roughly the same. All nodes have mini-batches assigned according to their computation resources. 
 We then discuss the key steps of the proposed framework.

 \textbf{Random committee-construction.} All nodes would have a random identity. According to the identities, committees of size $c$ are formed. Every committee member knows the identity of all honest peers with high probability. After a certain number of rounds of training, a portion of nodes would be replaced by reliable nodes to prevent slowly-adaptive adversary\cite{kokoris2018omniledger}.

 \textbf{Intra-committee consensus.} In the first round, every member gossips its local gradient with the timestamp of the gossiping among its committee. In every step, a leader would aggregate two things: an agreed selection of $c^2/n$ gradients and the neighbor’s aggregation from last step. Meanwhile, members of the committee are required to validate the result. After a global stabilization time (GST), an agreement is reached by a partially synchronous consensus protocol Hotstuff \cite{yin2018hotstuff} in every committee. With a sufficient number of members having agreed and partially signed (threshold signature) on the data, the data is tamper-proof. Only the partial aggregation result is to be transmitted to the neighbor committee.

 \textbf{Global consensus.} Similar to the Ring Allreduce\footnote{\url{https://github.com/baidu-research/baidu-allreduce}} process, committees communicate locally agreed data with their neighbor committees. After $2(n/c - 1)$ steps, every node in the system would finally have a globally agreed aggregation.

 \textbf{Reconfiguration.} After some iterations of training, some malicious nodes in the committees are replaced. We use Cuckoo Rule \cite{zamani2018rapidchain} against join/leave attack while maintaining a same resiliency of 1/3 after reconfiguration. 


\subsection{Intra-Committee Consensus}

 As illustrated in figure \ref{fig:Process}, a learning task proceeds in \emph{iterations}, wherein each iterations consist of \emph{steps}. In an iteration, all nodes start by computing their local gradients and ends by updating the model parameters as in all distributed learning procedures. {\color{black} In each step, all nodes take part in a consensus process to ensure that byzantine efforts are limited.}

 \begin{figure*}[t]
  \centering
  \includegraphics[width=1\linewidth]{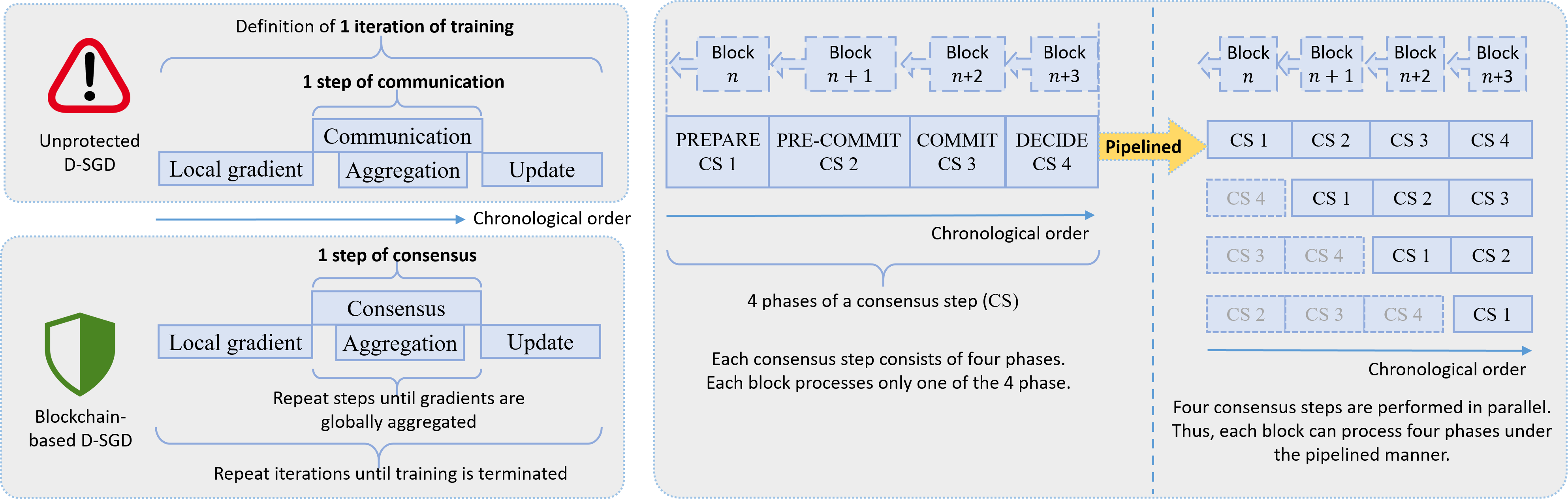}
  \caption{We define the process of finishing a partial aggregation a \emph{step}, the process of finishing a model update an \emph{iteration}. A consensus step (CS) has 4 phases, each driven by leader. The protocol can be pipelined for performance enhancement, ideally executing 1 aggregation each block in average.}\label{fig:Process}
\end{figure*}

 When a local gradient is computed, members take part in repeated consensus steps. 
 A consensus step (CS) includes:
 
  \begin{itemize}
    \item \textbf{Component 1}: local gradient selection, 
   
    \item \textbf{Component 2}: neighbor committee aggregation, 
   
    \item \textbf{Component 3}: aggregation result.
    \end{itemize}
    
    Fig. \ref{fig:Process} depicts the intra-committee consensus process. We define the process of finishing a partial aggregation a \emph{step}, the process of finishing a model update an \emph{iteration}. We utilize the three-phase chained Hotstuff consensus protocol\cite{yin2018hotstuff}, wherein a consensus step require 4 phases of communication.  Each phase is driven by a leader issuing a block containing verifiable quorum certificates. For component 1, committee members can either collaboratively select $c^2/n$ local gradients, or coordinate in a round robin manner to choose $c^2/n$ local gradients. Having a neighbor committee aggregation and a set of local gradients, both leaders and members can aggregate them using the detection-based BFT aggregation \cite{li2019abnormal}. A pre-trained anomaly detection model would assign a weight to each gradient according to the anomaly score. If the anomaly score surpasses a threshold, zero weight would be assigned accordingly, thereby “filtering” harmful gradients that hinders convergence. Meanwhile, members are required to store historical credit scores of each other until a new committee is formed. Before the committee reconfiguration, credit scores are transmitted to the permission control center.

 For component 2, members wait for the leader of the neighbor committee to broadcast the tamper-proof neighbor aggregation from last step. Since component 2 is an agreed aggregation result from the neighbor’s last step, it can be easily verified. Finally for component 3, the incumbent leader broadcast the partial aggregation and a hash index of training parameters for members to verify and agree on. Having sufficient quorum certificates, the leader would broadcast the decided aggregation in the committee and in the neighbor committee. If the leader chooses to withhold the result from the neighbor committee, the neighbor committee can ask a random committee member for the result.

 As shown in Fig. \ref{fig:Process}, in order to reach a consensus decision (CS4), three phases of validation are required. {\color{black}This means that only 1/4 of blocks are generating aggregations}. To address this issue, we can pipeline consensus steps as encapsulated commands to achieve a better performance. As Fig. \ref{fig:Process} shows, utilizing the frequent view-change of Hotstuff, each leader would be responsible for driving four consensus steps in different phases, with each phase validating an aggregation proposal. In an ideal scenario where no byzantine leader is elected, every block generates an agreed aggregation in average. Members are required to store four sets of gradients for validation, with each set composed of its own local gradient, corresponding aggregation of neighbor committee and an aggregation proposal from incumbent leader. {\color{black} Since PIRATE does not rely on rollback to recover from misdirection of the byzantine leadership, computing nodes are free from the storage burden of the entire history of all gradients}.


\section{Case Study}

\subsection{Implementation and Methods}

We implement a prototype of PIRATE with python 3.7 for performance evaluation. On one machine, we run 50 to 100 instances each to simulate a single committee. Assuming after permission control, devices spend a same amount of time for computing gradients. We {\color{black} simulate this process by having instances wait for a same amount of time to generate an equal-sized chunk of data (28MB)}. Then, instances transmit chunks of data to simulate a decentralized SGD process. {\color{black} The machine is a mini PC (model serial number: NUC8i5BEK), with a quad-core i5-8259U processor (up to 3.80 GHz). To simulate the network condition of 5G, we assume every message has a 10ms latency. And the uplink bandwidth is uniformly-distributed ranging from 80Mbps to 240Mbps, while the downlink bandwidth is set to 1Gbps for each node.}

In order to minimize the communication overhead of gradient broadcast, we fix the ratio between $n / c^2$ to $4/1$, to ensure exactly one local gradient and neighbor’s gradient is aggregated in each consensus step.

\subsection{Performance Evaluation}

We compare our prototype with another blockchain-based SGD framework LearningChain without the presence of malicious node. We first compare the gradient storage overhead of the two frameworks. We then compare the iteration time measured by the time used to broadcast a block that contains the necessary information for nodes to progress in D-SGD.

As Fig. \ref{fig:Experiment} shows, gradient storage overhead for PIRATE nodes are constant as iteration progresses, while LearningChain’s storage has a linear growth. In each iteration, PIRATE stores only the leader’s gradient, the neighbor committee’s gradient and the local gradient itself has computed. In LearningChain, nodes are required to store the history of all leader-announced gradients, and the local gradients broadcasted by all nodes. 

Fig. \ref{fig:Experiment} shows that PIRATE outperforms LearningChain on iteration time, i.e., the time used for each node to get a model update. The major cost in each iteration is the broadcast of gradients. Broadcasting to a larger number of nodes significantly increase the cost. PIRATE shows a superior performance in terms of iteration time for each committee. {\color{black} This is because nodes are required to broadcast to only $c$ members, meanwhile, consensus decisions are reached in a parallel manner.}

\begin{figure}[t]
  \centering
  \includegraphics[width=1\linewidth]{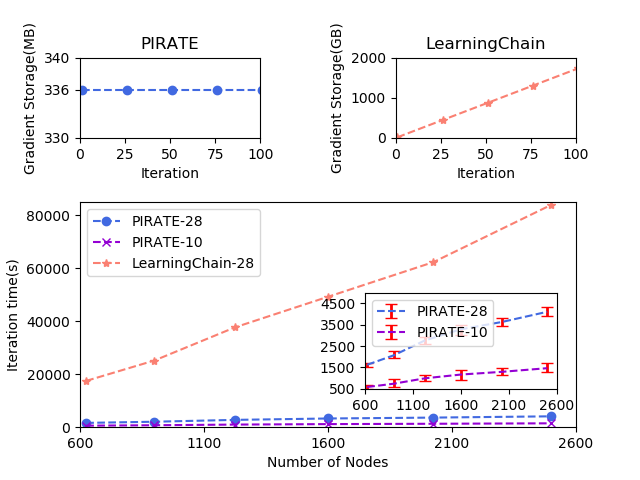}
  \caption{The left-top figure: Gradient Storage of PIRATE vs. iteration with single gradient size of 28MB. The right-top figure: Gradient Storage of LearningChain vs. iteration with the single-gradient size of 28MB. The Bottom figure: Iteration time of PIRATE and LearningChain vs. number of nodes with the single-gradient sizes of 28MB and 10MB.}\label{fig:Experiment}
\end{figure}

\section{Open Issues}

The open issues are envisioned as follows.

\begin{itemize}
    \item \textbf{Decentralized Permission Control}. Other than join/leave Sybil attacks, candidates having inferior computation ability, bad historical credit scores and unstable network conditions can undermine the efficiency of distributed learning. Without a centralized permission control, reliability assessment is challenging, especially for realtime states. However, for a decentralized system, latency induced by communication and verification inevitably affects timeliness.
    We can utilize an anchor chain to manage non-realtime states of devices to achieve further decentralization. The anchor chain is maintained by all candidates. Permissions to join shard chains, (i.e. learning committees) are jointly managed by the anchor Chain and a real-time assessment component.
    
    \item \textbf{Protection Against Model Poisoning Attack}. For federated learning, model-poisoning attack is a challenging issue in terms of safety. The attack can be successful even for a \emph{highly constrained} byzantine node \cite{bhagoji2018analyzing}. By exploiting the non-i.i.d property of data shards, byzantine attackers can send poisoned local updates that do not hurt convergence. Such harmful local updates can still affect the global model that triggers misclassifying. The attack method is also ``sneaky" enough to bypass accuracy check by a central server.
    In a decentralized environment, where nodes are less constrained in terms of communicating, computing and validating, model poisoning attack can be even more threatening.

    \item \textbf{Incentive Mechanism}. In a decentralized setting, each node shares greater responsibility than the nodes in a centralized setting. This is because they are constantly required to conduct actions, like validations and broadcasting, which seem non-rewarding through the perspective of an individual. 
    Since PIRATE is built based on a blockchain system, incentive mechanisms of participating can be naturally guaranteed. 
    
\end{itemize}

\section{Conclusion and Future Work}
To guarantee the high availability of distributed learning in 5G era, a distributed-learning framework with high efficiency, decentralization and byzantine-resiliency is in urgent need.
To fill this gap, we propose PIRATE, a byzantine-resilient D-SGD framework under the decentralized settings. Utilizing a sharding-based blockchain protocol, model parameters and gradient aggregations can be well protected. We then evaluate the feasibility of the PIRATE in a scenario where the large scale participants are with large-sized local models. Simulation results show that the proposed PIRATE is more efficient than an existing LearningChain solution in terms of both communication time and storage complexity.

We will further analyze the robustness and resiliency of PIRATE by conducting extensive experiments. Currently, we are  developing a byzantine-resilient aggregation algorithm dedicated for PIRATE, by considering the efficiency of computation and communication.

\section{Acknowledgement}
This work is partially supported by National Natural Science Foundation of China (61902445, 61872310), partially by Fundamental Research Funds for the Central Universities of China under grant No. 19lgpy222, and partially by Natural Science Foundation of Guangdong Province of China under Grant 2019A1515011798.

\bibliographystyle{IEEEtran}
\bibliography{Reference}

\begin{IEEEbiographynophoto}
{Sicong~Zhou}
(chowsch2@mail2.sysu.edu.cn) is currently with the School of Data and Computer Science, Sun Yat-Sen University, China. His research interests mainly include distributed learning and blockchains optimization.
\end{IEEEbiographynophoto}

\begin{IEEEbiographynophoto}
{Huawei~Huang} (M'16)
(corresponding author, huanghw28@mail.sysu.edu .cn) received his Ph.D in Computer Science and Engineering from the University of Aizu, Japan. He is currently an associate professor with the School of Data and Computer Science, Sun Yat-Sen University, China. His research interests mainly include distributed learning and blockchains optimization. He serves as a visiting scholar with the Hong Kong Polytechnic University (2017-2018); a post-doctoral research fellow of JSPS (2016-2018); an assistant professor with Kyoto University, Japan (2018-2019). He is a member of ACM.
\end{IEEEbiographynophoto}

\begin{IEEEbiographynophoto}
{Wuhui~Chen}
(chenwuh@mail.sysu.edu.cn) is an associate professor in Sun Yat-sen University, China. He received his bachelor's degree from Northeast University, China, in 2008. He received his master's and Ph.D. degrees from University of Aizu, Japan, in 2011 and 2014, respectively. From 2014 to 2016, he was a JSPS research fellow in Japan. From 2016 to 2017, he was a researcher in University of Aizu, Japan. His research interests include Edge/Cloud Computing, Cloud Robotics, and Blockchain.
\end{IEEEbiographynophoto}

\begin{IEEEbiographynophoto}
{Zibin~Zheng} (SM’16) 
(zhzibin@mail.sysu.edu.cn) received the Ph.D. degree from the Chinese University of Hong Kong, Hong Kong, in 2012.,He is a Professor with the School of Data and Computer Science, Sun Yat-sen University, Guangzhou, China. His current research interests include service computing and cloud computing.,Prof. Zheng was a recipient of the Outstanding Ph.D. Dissertation Award of the Chinese University of Hong Kong in 2012, the ACM SIGSOFT Distinguished Paper Award at ICSE in 2010, the Best Student Paper Award at ICWS2010, and the IBM Ph.D. Fellowship Award in 2010. He served as a PC member for IEEE CLOUD, ICWS, SCC, ICSOC, and SOSE. 
\end{IEEEbiographynophoto}

\begin{IEEEbiographynophoto}
{Song~Guo}
(M'02-SM'11-F'20) (song.guo@polyu.edu.hk) received his Ph.D. degree in computer science from the University of Ottawa, Canada. He is currently a full professor at Department of Computing, The Hong Kong Polytechnic University. His research interests mainly include cloud and green computing, big data, and cyber-physical systems. He serves as an Editor of several journals, including IEEE TPDS, TETC, TGCN, and IEEE Communications Magazine. He is a senior member of IEEE and ACM, and an IEEE Communications Society Distinguished Lecturer.
\end{IEEEbiographynophoto}

\end{document}